\documentclass{aa}
\thesaurus{} 
\usepackage{graphics}
\usepackage{psfig}     
\begin{document}
\def \arcm{\hbox{$^\prime$}}
\def \arcs{\hbox{\arcm\hskip -0.1em\arcm$\;$}}

\def\EE#1{\times 10^{#1}}
\def\gcm{\rm ~g~cm^{-3}}
\def\cm3{\rm ~cm^{-3}}
\def\kms{\rm ~km~s^{-1}}
\def\cms{\rm ~cm~s^{-1}}
\def\ergs{\rm ~erg~s^{-1}}
\def\wl{~\lambda}
\def\wll{~\lambda\lambda}
\def\FeI{{\rm Fe\,I}}
\def\FeII{{\rm Fe\,II}}
\def\FeIII{{\rm Fe\,III}}
\def\Msun{~{\rm M}_\odot}
\def\Ti44{M(^{44}{\rm Ti})}

\def\lsim{\!\!\!\phantom{\le}\smash{\buildrel{}\over
  {\lower2.5dd\hbox{$\buildrel{\lower2dd\hbox{$\displaystyle<$}}\over
                               \sim$}}}\,\,}
\def\gsim{\!\!\!\phantom{\ge}\smash{\buildrel{}\over
  {\lower2.5dd\hbox{$\buildrel{\lower2dd\hbox{$\displaystyle>$}}\over
                               \sim$}}}\,\,}

\title{The nature of the prompt X-ray and radio emission from SN 2002ap
\thanks{Based on
observations made with XMM-Newton Observatory and the Giant Metre
Wave Radio Telescope}}

\author{
F. K. Sutaria\inst{1}
\and P. Chandra\inst{2,} \inst{3} 
\and S. Bhatnagar\inst{4,} \inst{5}
\and A. Ray\inst{2} 
}
\institute{
Department of Physics and Astronomy, The Open University, Milton Keynes, U.K.
\and Tata Institute of Fundamental Research, Bombay 400 005, India
\and Joint Astronomy Programme, Indian Institute of Science, Bangalore 560 012, India
\and National Centre for Radio Astrophysics, Pune 411 007, India
\and National Radio Astronomy Observatory, Socorro, NM 87801, USA
}

\date{Received:  Accepted: }

\mail{akr@tifr.res.in}

\titlerunning{X-ray and radio emission from SN 2002ap}

\maketitle

\begin{abstract}

We report on the combined X-ray and radio observations of the type Ic
SN 2002ap, using XMM-Newton ToO observation of M74 and the Giant
Metrewave Radio Telescope (GMRT). We account for the 
presence of a nearby source in the pre-supernova Chandra field of view
in our measurements of the X-ray flux (0.3 - 10 KeV) 5.2 days
after the explosion. The X-ray spectrum is well fitted by a
power law spectrum with photon index $\alpha= 2.6$. Our results 
suggest that the prompt X-ray emission originates from 
inverse Compton scattering of photospheric thermal emission 
by energetic electrons. Radio observations with the GMRT at 610 MHz
(8 days after the explosion) and 1420 MHz (70 days after the explosion)
are combined with the high frequency VLA observations
of SN 2002ap reported by \cite{Ber02}, 
and the early radiospheric properties of SN 2002ap are compared 
with similar data from two other supernovae. Finally, 
the GMRT radio map reveals 
four other X-ray 
sources in the field of view of M74 with radio counterparts.

\end{abstract}



\section{Introduction}
\label{sec: Intro}
Supernovae are known to be explosions of massive and intermediate mass stars,
although the nature of the progenitor star for different supernova types,
remains an area of long-standing research.  Currently, it is believed that type
II and type Ib or Ic supernovae arise from core collapse of massive stars, while
the more homogeneous type Ia supernovae are the results of thermonuclear
explosions.  There is a considerable variety among the core collapse supernovae
spectroscopically and in their kinetic energy.  The `hypernova' class among
these are believed to have explosion energy significantly in excess of $10^{51}
\rm erg$, found in "normal" supernovae. The spatial and temporal 
near-coincidence of type Ic SN 1998bw with the long duration GRB980425 has
provided further impetus to observational and theoretical studies of type Ib/Ic
SNe and their progenitors (\cite{Mac01}, \cite{Mes93}).

%

During a supernova explosion the interaction of the outer parts of the stellar
ejecta with the circumstellar matter gives rise to a high energy density shell.
X-ray emission is expected from both shocked circumstellar matter and the
shocked supernova matter (see e.g. \cite{Che82}, \cite{Che01}).  
In addition, the interaction region may also accelerate electrons to
relativistic energies and amplify pre-existing magnetic fields
which gives rise to nonthermal synchrotron emission seen in many
supernovae.  
The radio and the X-ray emission give information on a
region of the supernova which may be far removed from the optical photosphere
(which has a smaller radius), although the conditions in the optical photosphere may
determine the X-ray emission characteristics in some instances.  In general,
X-ray and radio observations in early stages of a supernova can be used to determine
(1) the total mass lost from the pre-supernova star before explosion 
and (2) constrain
various physical processes leading to X-ray and radio emission.  

The type Ic SN 2002ap was discovered on Jan.  29.4, 2002 (Y.  Hirose as 
reported by \cite{Nak02} ) in NGC  628 (M74), at a distance of only $7.3$ Mpc.  
Based on spectral analysis 
of the early observations, the epoch of explosion was estimated at 
Jan.  $28.0 \pm 0.5$ 2002UT (\cite{Maz02}). For the purpose of the 
analysis presented in this paper, we will regard Jan.  $28.0$ as the date of 
explosion.  The broad spectral features (\cite{Kin02},  \cite{Mei02},
\cite{Gal02}), and a subsequent modeling of its spectroscopic and photometric 
data (\cite{Maz02}), suggested that this was an energetic event with
explosion energy $E \simeq 4 -10 \times 10^{51}$ erg. 
 In this paper, we discuss our analysis of the XMM-Newton observation in 
the X-ray (0.1-15 keV) bands  (Sect. \ref{sec: XrayObs}), accounting for the 
presence of a nearby object in the pre-supernova X-ray field, observed 
earlier by the Chandra X-ray Observatory (Sect.  \ref{sec: Chandra}). We also
observed SN 2002ap in the 0.61 GHz and 1.4 GHz radio bands 
(see Sect. \ref{sec: GMRT}), and the implications of the  GMRT upper 
limits in the context of VLA observations at the same epoch (\cite{Ber02})
are discussed in Sect. \ref{sec: combined}. 

We also summarize the explosion parameters that we 
derived from the optical observations and modeling reported by \cite{Maz02},
as this is used as input for later sections (Sect. \ref{sec: combined}).  
We have also combined the GMRT
data with the VLA data (\cite{Ber02}) to derive conditions near the
radiosphere. A combined analysis of the early X-ray and radio observations is
presented in Sect. \ref{sec:  massloss}, which attempts to constrain the
multiple physical processes (thermal and non-thermal) that are responsible for
the early X-ray emission. We also compare SN 2002ap with 2 other SNe (1998bw (Ic) 
and 1993J (IIb) for which early multi frequency data is available. 
Finally, in Sect. \ref{sec:  radiative} we discuss
these results in the context of the presupernova star and its evolution.

\section{X-ray observations and data analysis.} 
\label{sec:  XrayObs}
 
SN 2002ap and its host galaxy M74 (NGC 628) were observed by XMM-Newton,
four days after the discovery of the SN. 
\cite{Pas02}  and \cite{Sor02} have reported the EPIC-PN detection of
the supernova, as well as several other sources in the host galaxy.
Here we present a spectral analysis of the XMM-Newton EPIC-PN and -MOS
CCD detection, taking into account contributions from other
pre-existing sources present in the extraction circle, near the supernova.

XMM-Newton observed the field of view of SN 2002ap in the full-frame,
thin filter mode,  for the EPIC-PN and the two EPIC-MOS
cameras. Simultaneously, it also observed M74 with the Optical
Monitor in the UVW1. The full exposure in the
EPIC CCDs was $\sim 34$ ks between Feb 3.00 to 3.40, 2002 UT.
Since Chandra X-ray observatory  also observed the same field on
June 19, 2001 and October 19, 2001 for a total of 47 ks, we
present our analysis, based on both the presupernova and post-supernova
exposures.

Early analysis of the XMM-ToO observations of SN 2002ap showed that the 
supernova is rather faint in X-rays with a flux of $\sim 10^{-14}$ erg 
cm$^{-2}$ s$^{-1}$. 
Because the psf of the XMM EPIC-CCDs is rather large ($\sim 50 \arcs$ to 
enclose 90\% of the energy at 10 keV), it was necessary to verify the 
absence of, or to account for the presence of any other sources, however 
faint, within the spectral extraction region on the EPIC CCDs. 

\subsection{Chandra observations of the presupernova field:}
\label{sec: Chandra}

In order to check if there were any other sources present in the vicinity
of the optical position of SN 2002ap (RA $= 01^{h} 36^{m} 23\fs85$, 
Dec = $+15\degr 45\arcmin 13\farcs 2 $), we examined the archival
Chandra observations of M74, 
carried out on 19th Oct. 2001. In the net 46.2 ksec of exposure time with 
Chandra-ACIS, we did not see any source at the position of the supernova. 
However, we did find a bright source at RA $01^{h} 36^{m} 23\fs404$ and 
DEC $ +15\degr 44\arcm 59\farcs89$, which, being only 14.9 \arcs away
from the SNe, was well within the region of 
XMM spectral extraction. Comparing the positions of bright, point-like sources 
seen in both the Chandra and XMM fields of view, we found that the relative 
astrometric shift between 
Chandra and XMM is at most $\sim 5\arcs$. Thus, this object 
(hereafter, CXU J013623.4+154459) would
have been detected within the 40\arcs circle used to extract the XMM-EPIC 
spectra. 

Level-2 pipeline processed Chandra 
data was used for spectral analysis,  since
a check on the calibration files used in the archived level-2 processed data
showed that no improvement would be made by re-calibrating the data. The
spectrum was extracted using the CIAO software, 
 and the spectral analysis was carried out using the SHERPA software.
The source was located on ACIS-S6 chip, and a light curve was extracted  
from a source free region of the ACIS-S6 chip to
check that the observation was not contaminated by background flares. 
We also extracted a light curve for the source, to check 
for flaring, or any periodic variation, but the object is too faint 
for any such variability to be noted. Since no sources are seen within $10 
\arcs$ of this object, 
we used a region of radius $6 \arcs$ to extract the spectrum. 
As a measure of the PSF at the source location, a radial region of  
$\simeq 1.5 \arcs$ would enclose 50\% of the energy at 1.49 keV. 


CXU J013623.4+154459 is very faint, and the higher energy ($E \ge 9$ keV) 
spectrum is 
seen to be strongly dominated by the background --  possibly due to
high energy particles, rather than X-ray events. Because of the problems 
associated with accurately modeling the high energy component of the 
background spectra, we used background subtracted data and restricted 
our analysis to the energy range $0.3-10.0$ keV. The background subtracted 
count rate in this energy band is $6.5(\pm 2.4) \times 10^{-4}$ s$^{-1}$. 
Fixing the column density at $0.49 \times 10^{21}$ cm$^2$, 
(obtained from online hydrogen column density generator
of High Energy Astrophysics Archive (HEASARC), in the direction of the SN)
and using the $\chi^2$-Gehrels statistic, we found that the source was well 
fitted 
by a power-law spectrum with a photon index $0.6^{+0.6}_{-0.35}$, implying a 
0.3-10.0 keV flux of $1.3 \times 10^{-14}$ ergs cm$^{-2}$ s$^{-1}$. The 
$\chi^2$ goodness-of-fit parameter was $Q=0.899$, where $Q$ measures the
probability of obtaining the observed (or larger) value of $\chi^2$, if 
the data was well represented by the fitted parameters. For an ideal fit,
$Q=1$.

\subsection{XMM-Newton observations of SN 2002ap field:}
\label{sec: XMM}

SN 2002ap was observed by the XMM instruments from Feb. 2.03 to Feb. 2.42 UT, 
2002 (estimated 
5 days after the
explosion), for a duration of 37.4 ksec. Both EPIC-MOS and EPIC-PN 
observations were carried out in the 'Thin1' filter mode.
The data was pipeline processed and calibrated using the XMM-Science Analysis 
Software (SAS) version 5.2, 
and the latest available versions of the calibration files. 
The events in the EPIC-MOS1, 
EPIC-MOS2 and EPIC-PN data sets were filtered using the
SAS-xmmselect task, using the appropriate flags and event selection criteria to
account for event pile-up. We followed the analysis procedure recommended 
in ``Users Guide to the XMM-Newton Science Analysis System" (2001)), with 
the exception of allowing patterns $\le 4$ in the case of EPIC-PN, to account
for any pile-up, however negligible it may be.
A time series analysis of the entire EPIC-PN, MOS1 and MOS2 data, 
shows that there are times  of large, random fluctuation over the entire field 
of view of both EPIC-PN and MOS CCDs. Filtering out these intervals of highly 
fluctuating background reduced the exposure time to 25.5 ks in EPIC-PN and 
$\sim 30$ ks in each of EPIC-MOS1 and MOS2.

SN 2002ap was identified on the EPIC-CCDs using the optical 
coordinates (RA $= 01^{h} 36^{m} 23\fs85$, DEC $=+15\degr 45 \arcmin 
13\farcs2$). The EPIC-PN spectra was extracted using a circle of radius 
$40 \arcs$, within which almost 80\% of the source 
photons in the energy range $10.5 - 1.5$ keV would be enclosed, with the
extraction region remaining on a single CCD chip. The EPIC-MOS1 and 
EPIC-MOS2 spectra were extracted using a psf of radius $50\arcs$, which
encloses 90\% of the total energy. The response matrices and ancillary 
response files were generated using the SAS-rmfgen and SAS-arfgen tasks, and
the appropriate line number for the PN-CCD. The data was grouped and analysed 
using the XSPEC software.  

\subsection{Results of X-ray observations}
\label{sec: Xray}

The 0.3-12 keV, background subtracted, count rate in XMM EPIC-PN is 
$6.6 (\pm 1.0) \times 10^{-3}$  s$^{-1}$. In the same energy range, 
the EPIC-MOS1 
count rate is $1.69 (\pm 0.64) \times 10^{-3}$ s$^{-1}$ and the EPIC-MOS2 
count rate is $1.7 (\pm 0.64) \times 10^{-3}$ s$^{-1}$.   

\begin{figure}
\resizebox{\hsize}{!}{\rotatebox{270}{\includegraphics*[4.0cm,1.7cm][20cm,25.1cm]{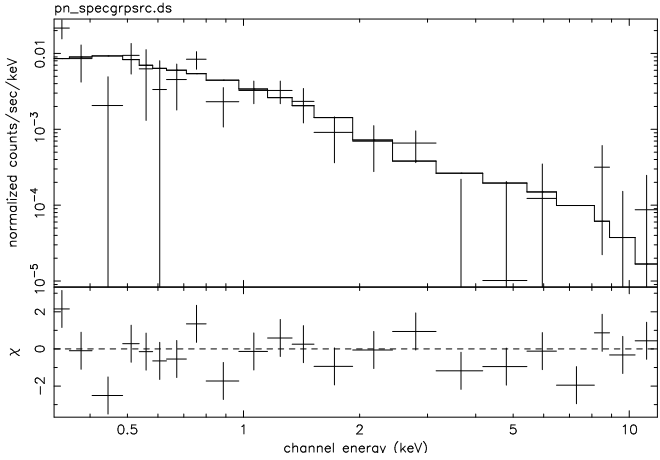}}}
\caption{ EPIC-PN spectrum and the fitted power law model 
(see Sect. \ref{sec: Xray}) of SN 2002ap.
The model fitted here consists of an absorbed powerlaw of photon 
index = -2.6 with an additional constant absorbed power law component 
to account for CXU J013623.4+154459. $\log N_H$ was fixed at 20.68.
The residuals are plotted in the lower panel.
}
  \label{fig: powerlaw}
\end{figure}

In order to account for the presence of the Chandra source discussed above 
in the XMM
spectral extraction region, all models fitted to the data had an additive
absorbed power-law component, with parameters derived from fits to the Chandra
data, as discussed above. Since the SN is very faint, and the spectra
rather sparse (Fig. \ref{fig: powerlaw}), 
we kept $N_H = 0.049 \times
10^{22}$ cm$^{-2}$ as constant for all fits. In general, we found that letting 
$N_H$ vary can often result in unphysically low values of 
$N_H \ll 10^{-16}$ cm$^{-2}$, while providing little improvement in the value
of reduced $\chi^2$. 
In Table \ref{tab: EPIC-PN}, we have quoted the parameters from the best
fitted models in our analysis  -- 
the 0.3-10 keV flux quoted there is corrected for the presence of the 
Chandra source. The data is rather sparse, hence $\chi^2$-fitting is unable to 
distinguish between thermal bremsstrahlung and the simple power law models.
The power law distribution (Fig. \ref{fig: powerlaw})
fits the data well with $\alpha = 2.6$, as does the
thermal bremsstrahlung model with temperature $kT \sim 0.8$ keV.
Using the powerlaw model, we find that the total flux in the EPIC-PN extraction
region is $ 2.4 \times 10^{-14}$ ergs cm$^{-2}$ s$^{-1}$. Correcting for the
presence of CXU J013623.4+154459, the 0.3-10.0 keV flux from 
from SN 2002ap is $1.07^{+0.63}_{-0.31} \times 10^{-14}$ 
ergs cm$^{-2}$ s$^{-1}$.

Adding an extra power-law, or a cutoff power-law component to the 
bremsstrahlung, or Raymond-Smith or blackbody models did not improve the fit, 
and resulted in poorly constrained values of temperatures with an 
unusually high 
photon index $ \alpha \sim 9.0$, and hence these models were discarded. 

\begin{table}
\caption{Best fit spectral parameters for EPIC-PN. $kT$ represents
the plasma temperature or the black body temperature, depending on the model.
All uncertainties are quoted in 90\% confidence limit. 
$N_H$ was constant for all
fits. The absorbed flux for SN 2002ap is quoted here.
 \label{tab:
EPIC-PN}}
\begin{tabular}{cccccc}
\hline
\hline
Model & $N_H$ & $\alpha$ & $kT$ & $\chi_{\nu}^2/dof$ & $f|_{0.3-10}$ \\
\hline
 & $10^{21}$ &  &  &  & $  10^{-14}$ erg \\
 & cm$^{-2}$ & & keV &  & /cm$^2$/s \\
\hline
Power-law & 0.49 & $2.6_{0.5}^{0.6}$ & --  & $ 1.2/20 $ & 1.07 \\

          & 0.42 & $2.5_{0.5}^{0.6}$ & --  & $ 1.2/20 $ & 1.0 \\

Thermal & 0.49 & -- & $0.84_{-0.3}^{+0.9}$ & $ 1.2/20$  &  0.81 \\
Brems. & & & & & \\ 

Raymond-  & 0.49 & -- & $2.31_{-0.8}^{+1.9}$ & $1.58/20$ & 1.04 \\
Smith     &       &    &                      &           &     \\

Blackbody & 0.49 & -- & $0.21_{-.06}^{+0.1}$ & $1.4/20$ & 0.6 \\

\hline
\end{tabular}
\end{table}

Because of the low count rate in EPIC-MOS CCDs, it was decided to fit to the
combined MOS1 and MOS2 data, allowing only the relative normalisations to vary.
As discussed for EPIC-PN data, each model incorporates an absorbed 
power-law component to account for the Chandra source. Thus, the data was 
well fitted by an absorbed thermal bremsstrahlung, with 
$\chi^2/d.o.f. = 0.95/19$ and $kT = 
0.4_{-0.22}^{1.28}$. The 0.3-10 keV flux (corrected for CXU J013623.4+154459
) was $7.1 \times 10^{-15}$ ergs cm$^{-2}$ s$^{-1}$.

SN 2002ap was too faint to be detected by either of the RGS detectors.   

\subsection{The Optical Monitor observation.} 
\label{sec:  OM}

The XMM-Newton Optical Monitor observed SN 2002ap in the 4-frame
(ENG-2) mode with UVW1 filter for a duration of 2.5 ks.	The UVW1
filter peaks at $ \lambda \simeq 270$ nm, and spans the 
range $\sim 250 \le \lambda \le \sim 310$ nm.
The frames were combined and the data was pipeline processed using SAS
version 5.3 with the latest available calibration files. A source list
was compiled. We find that the optical monitor count rate for SN 2002ap
was $14.67 (\pm 0.08)$ s$^{-1}$, with the background count rate being 
$0.8$  s$^{-1}$. This corresponds to a UVW1 flux of $7.667 (\pm 0.002)
\times 10^{-15}$ erg cm$^{-2}$ s$^{-1}$ ${\AA}^{-1}$.

\section{GMRT observations of SN 2002ap field}
\label{sec: GMRT}

The radio observations of SN 2002ap were made with the Giant Metrewave Radio
Telescope in aperture synthesis mode on 2002 February 5.56 in the 610
MHz band and on 2002 April 8.22 in the 1420 MHz band. 

For both the observations at 610 as well as 1420 MHz, the pointing centre
was chosen to be the position of supernova i.e.  RA= $ 01^{h} 36^{m} 23\fs85$
and Dec$=15\deg 45\arcm 13\farcs2$.  In 610 MHz observations the time on
source were around 2 hours and 3C48 was used as flux calibrator as well as
phase calibrator.  It was also used as bandpass calibrator. 3C48 was
observed for approximately 10 minutes after every 30 minutes.  3C48 
flux at 610 MHz was derived from the best fit spectra given in VLA calibrator manual.
Out of a total of 30 antennas, we had 22 good
antennas.  The full band of 16 MHz was used, resulting in 128 frequency
channels of width $\sim 125$ kHz.  For the observations in 1420 MHz band, the
number of antennas used were 24.  The time on the source was $\sim$ 3.5
hours.  3C48 was used as a flux calibrator and it was observed at the
beginning and at the end of the observations.  The flux of 3C48 was assumed
to be 16.15 Jy from VLA calibrator manual.  Compact sources 0119+321
and 0238+166 were used as the phase
calibrator with fluxes 2.6 Jy and 1.26 Jy respectively.  3C48 was also used
as a bandpass calibrator.  The full band of 16MHz resulting in 128 channels
of width $\sim$ 125 kHz was used.  

\subsection{Data reduction, calibration and results}
\label{sec: radioresults}
The data was analysed using AIPS (Astronomical Image Processing Software) of
National Radio Astronomy Observatory. Interference in some of the frequency
channels, spikes and dropouts due to electronics etc. were flagged using
the AIPS task FLGIT. FLGIT was run on the calibrator scans and 10 times
the expected noise per channel was used as the flagging criterion. 
In this manner
15\% of the data was flagged. 
Then one clean channel was looked at and antenna and baseline
based flagging was done. Approximately 20\% 
data was lost in the flagging process. After computing  
the single channel antenna-based complex gains for the flux calibrator 
and phase
calibrator with the time resolution of 20 seconds, bandpass calibration
was done using 3C48 to apply the single channel calibration to all the channels.
To get rid of the effect of the band smearing at low frequency (610 MHz), a
pseudo-continuum data base of 6 frequency channels was made from the central
100 channels of 16 MHz bandwidth.
Field maps were made by Fourier transformation and cleaning.
Iterative self-calibration rounds were run for phase
correction. Details of GMRT data analysis procedures are given in (\cite{San00}).

\begin{table}
\caption{GMRT observations for Flux and Phase Calibrators}
\begin{tabular}{cccccc}
\\
\hline

Source  & Date in     & RA         & Dec.    & $\nu$        & Flux \\
        & 2002   & J2000      & J2000   & MHz              & Jy \\
\hline

3C48    & 5 Feb  & 01 37 41.3 & 33 09 35 &610            & 29.4 \\
3C48    & 8 Apr  & 01 37 41.3 & 33 09 35 &1420            & 16.2 \\
0119+321& 8 Apr  & 01 19 35.0 & 32 10 50 &1420            & 2.9  \\
0238+166& 8 Apr  & 02 38 38.9 & 16 36 59 &1420            & 1.0  \\

\hline
\end{tabular}
\end{table}

\begin{figure}
\resizebox*{\hsize}{!}{\includegraphics{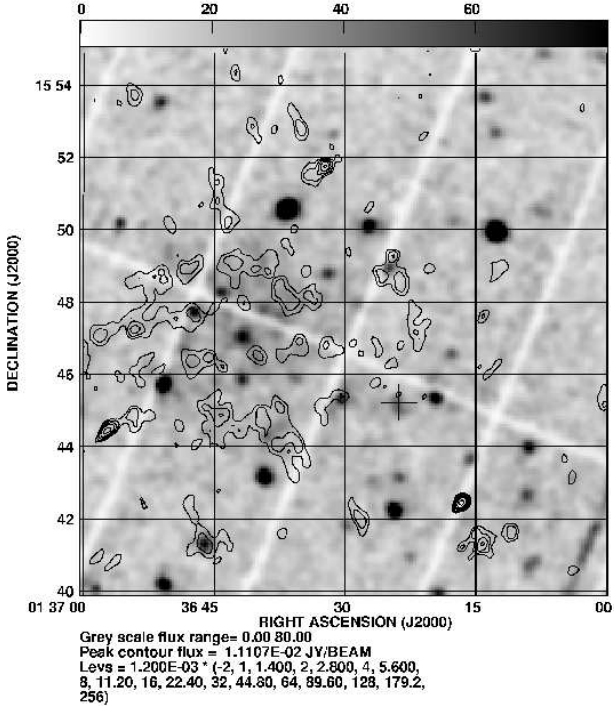}}
\caption{GMRT image of the field at 610 MHz containing SN 2002ap
overlayed on the X-Ray image from XMM. X-ray image is on a grey scale
while  the radio image is the contour map with lowest level of 1.2 mJy.
The image resolution is $15 \arcs \times 15 \arcs$. Four X-ray
sources having radio counterparts  are seen in this figure.
They are numbered according to their decreasing RA in Table
\ref{tab: othersources}. Position of the SN on the map is marked with
a cross.}
\label{fig:  combined}
\end{figure}
In neither of the observations carried out on 2002 February 5
(day 8.56) and on 2002 April 8 (day 42), SN 2002ap was detected by
GMRT. The upper limits of the fluxes from the region of the SN
are given in Table 3. The
GMRT 610 MHz contour map shows several extended sources, which are
likely associated with the spiral arms of the galaxy. Superposition
of the 610 MHz contour map on the X-ray EPIC-PN image 
(Fig. \ref{fig:  combined}), reveals the existence of 4 X-ray sources 
with radio counterparts. These are listed in Table \ref{tab:
othersources} with their radio flux densities and X-ray 0.3-10.0 keV 
count rates. Spectral analysis to determine the nature of these
sources is being done separately.


\begin{table}
\caption{Results of GMRT observations of SN 2002ap}
\begin{tabular}{ccccc}
\\
\hline

Date in       & $\nu$ & Resolution    & 2$\sigma$ Flux &  RMS  \\
2002     & (MHz)     & (arcsec)      & (mJy)  & mJy/ beam \\
\hline

5Feb    & 610       &9.5 x 6        & $< 0.34$         & 0.17 \\
8Apr    & 1420      & 8 x 3         & $< 0.18$         & 0.09 \\
\hline
\end{tabular}
\end{table}

\begin{table}
\caption{List of X-ray sources with radio counterparts in M74.
\label{tab: othersources}}
\begin{tabular}{ccccc}
\\
\hline
Source  & RA            & Dec.          & Radio         & EPIC-PN \\
No.  &               &               &    Flux                & count-rate \\
        & J2000         & J2000         & (mJy)              & $10^{-3}$ ct s$^{-1}$ \\
\hline
1       & 01 36 47.2  & 15 47 45  	& $7.1 (\pm 1.3)$    &  $4.3 (\pm 1.3)$ \\
2       & 01 36 46.1  & 15 41 17  	& $12.8 (\pm 1.5)$   &  $3.9 (\pm 1.3)$ \\
3       & 01 36 24.9  & 15 48 58  	& $22.7 (\pm 1.5)$   &  $1.9 (\pm 1.3)$ \\
4       & 01 36 30.5  & 15 45 17  	& $4.5 (\pm 1.3)$    &  $4.6 (\pm 1.3)$ \\
\hline
\end{tabular}
\end{table}

\section{The location of the radiosphere}
\label{sec: combined}



The radio turn-on of supernovae are found to be wavelength dependent, with
the flux at shorter wavelengths peaking before that at longer
wavelengths (e.g. \cite{Ber02}). This decreasing absorption can be either
via free-free absorption (FFA) or  synchrotron self-absorption (SSA) in the
expanding CSM. 
We have fitted the VLA and GMRT data on 2002 Feb 5.96 to a 
 SSA model, with spectral index = -0.8 in the optically thin
limit, implying that the radius of the
radio photosphere on this day  was $R_r = 3.5 \times
10^{15} \rm cm$ and the magnetic field in the shocked ejecta 
was $ B = 0.29 G $
(see Table \ref{tab: bestfit8.96}). 
We also note that that SSA prediction of flux at 610 MHz band is
consistent with the GMRT upper limit.

In order to compare the photospheric radii across wavebands, we
also compute the size of the optical photosphere at the same epoch. 
Using the synthetic light curve best-fitted to the bolometric light curve on 
day 5 (\cite{Maz02}), the bolometric magnitude was $M_{bol}= -16.5$, and the
visual magnitude was $ M_v= -17.1$ (using a distance modulus of
29.5 (\cite{Sha96})). This implies that the 
radius of the optical photosphere on day 5 was $R_{opt}=
3.4 \times 10^{14} \rm cm$, and corresponding total flux
across UV, optical and IR bands was 
$F_{UVOIR} = 2 \times 10^{-10} \rm erg \; cm^{-2} \; s^{-1}$. The inferred
conditions in the optical photosphere are listed in Table \ref{tab: optical}.


In contrast, the position of the radio photosphere on day 7 was
$r = 4.5 \times 10^{15} \rm cm$ and its velocity
$90,000 \; \rm km \; s^{-1}$
(\cite{Ber02}). Scaling to day 5, $R_{\rm radio} \sim 3.2 \times 10^{15} \rm cm$.
Thus, the radio photosphere is quite far outside the optical
photosphere. 
The region of production of the X-rays is also outside the
optical photosphere radius (see Sect. \ref{sec: radiative}). 
This substantiates the inference from
the  early spectrum (\cite{Mei02}) that there was sufficient
material at $ v \ge 30,000 \;\rm km\; s^{-1}$.

\begin{table}
\caption{Optical photosphere of SN 2002ap on day 5 (Feb 2, 2002). 
The distance modulus was $\mu = 29.5$, and the maximum value
of bolometric correction was used (\cite{All76}). 
\label{tab: optical}}
\begin{tabular}{cccccc}
\\
\hline
$M_{v}$   & $M_{bol}$        & $T_{eff}$       & $R_{opt}$   &$\bar{v}_{ph}$     & $F_{UVOIR}$ \\
          &                 & K               & cm.             & km/s              & $\rm erg cm^{-2} s^{-1}$\\ 
\hline
-17.4   & -16.5         & 11000         & $3.4\times 10^{14}$  & 8000       & $2\times 10^{-10}$\\
\hline
\end{tabular}
\end{table}

The framework of spherical symmetry has been used to perform the above
analysis. This is a reasonable assumption on the day of the XMM observation
and shortly afterwards, since the month-long, early
spectropolarimetric data of \cite{Wan02} implied there was little distortion of
the photosphere on 2002 February 3.

\begin{table}
\hspace*{5.5cm}
\caption{Best fit parameters of SSA for SN 2002ap on day 8.96 i.e. Feb 5.96, 2002
\label{tab: bestfit8.96}}
\hspace*{1.5cm}
\begin{tabular}{ccccc}
\\
\hline
$\alpha$   & $\nu_{p}$        & $F_{p}$       & $R_{r}$        & B    \\
         &GHz             & $\mu$Jy      & cm.          &  G    \\
 
\hline
0.8     & 2.45           & 397          & $3.5\times 10^{15}$ & 0.29 \\
\hline
\end{tabular}
\end{table}

Finally, we compare the early radiospheric properties
of SN 2002ap with those of other radio SNe, across type
classification, using equipartition arguments.  
Theoretically, early radio
emission from SNe is thought to originate from an envelope that is expanding
at a substantially higher speed than the optical photosphere
(\cite{Shk85}). 
It is only at early times after the explosion that the effect of 
Compton boosting by electrons to high energy bands producing
nonthermal spectrum is potentially observable (Sect. \ref{sec: radiative}). 
At such early times radio spectra of supernovae are relatively rare.

\begin{figure}
\resizebox*{\hsize}{!}{\includegraphics{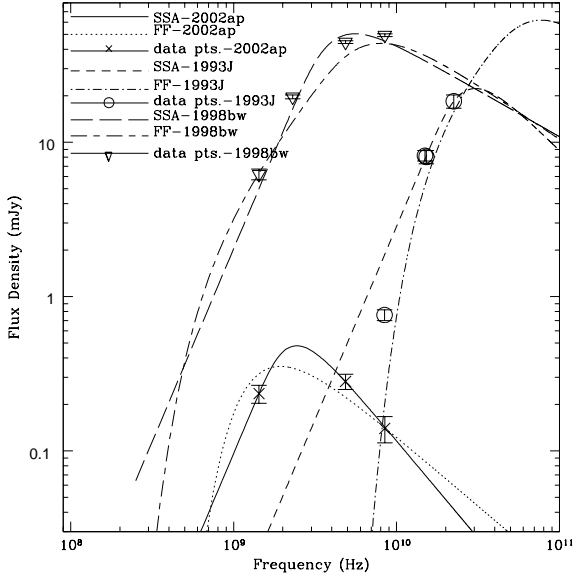}}
\caption{Comparison of spectrum of three SNe (SN1993J (\cite{vanDyk94}) ,
SN1998bw (\cite{Kul98}) and SN 2002ap (\cite{Ber02}) )
near day 11 after explosion.
Solid line, dashed line and long dashed lines show the best fit synchrotron 
self absorption model  for SN 2002ap, 1993J and 1998bw respectively and 
dotted line, dot-dashed line and short-long dashed line show the corresponding 
free free absorption model.}
\label{fig: 3SNe}
\end{figure}

We have fitted models (Fig. \ref{fig: 3SNe}) to the radio spectra obtained 
from three supernovae:  SN 1993J (a type IIb SN) on day 11.45,
SN 1998bw (a type Ic SN) on day 11.7 and SN 2002ap (another type Ic SN) 
on day 11.0. Best fit parameter values for SSA model were obtained
by using a $\chi^{2}$ fit.
While SN 1993J and SN 1998bw were much brighter at their
peaks compared to SN 2002ap, the synchrotron self-absorption peak for
SN 2002ap seems to occur at the lowest frequency among all three.
Best fit parameters  values are listed in Table \ref{tab: threesne}. 
\begin{table}
\caption{ SN 1993J, SN 1998bw and SN 2002ap best fit parameters of SSA on day 11
\label{tab: threesne}}
\small
\begin{tabular}{ccccccc}
\hline
SNe    &$\nu_{p}$ & $F_{p}$ & $\Theta_{eq}$ & $U_{eq}$             & $B_{0}$ & $R_{0}$             \\
       &          &         &               & $\times 10^{45}$                    &
& $\times 10^{15}$ \\ 
       &  GHz     & mJy     & $\mu$as      & erg                 & G       & cm                  \\
\hline
2002ap & 2.45     & 0.48    & 39.0          & 0.69  & 0.47    & 4.80 \\
1998bw & 5.5      & 50.4    & 112.4         & 3500   & 0.23    & 68.4 \\
1993J  & 30.5     & 22.3    & 17.6          & 0.50  & 3.54    & 1.08 \\
\hline
\end{tabular}
\end{table}
The equipartition (angular) radius (that gives the minimum
energy density in combined magnetic and relativistic particle energy 
densities) obtained by  \cite{Sly63}:

$$\theta_{\rm eq} 
= 120 d_{\rm Mpc}^{-1/17} S_{\rm p,mJy}^{8/17} \nu_{\rm p,GHz}^{(-2\alpha -35)/34} \rm \mu as$$ 
appears to be the largest for SN 1998bw (112 $\mu$as)
while those for SN 1993J and SN 2002ap are comparable (18 and 39 $\mu$as
respectively).  Hence SN 1998bw which was not only coincident with a GRB
source but also very bright in the radio was rapidly expanding. It 
was unusual compared to the more
ordinary SN like SN 2002ap (despite both supernova's identical type
classification), -- the latter having similarity with SN1993J in terms of
equipartition size, across type classification.

\section{Radio \& X-ray turn-on \& progenitor mass loss}
\label{sec: massloss}

The early emission in X-ray and Radio bands allows us to
estimate the density of the CSM along line of sight, and hence the mass loss 
rate from the presupernova star, if the terminal wind speed 
of the progenitor star is known or estimated.

%
%

The CSM density is related to the mass loss parameters as:
$ \rho_{\rm cs} = \dot M / 4 \pi u_{\rm w} r^2 $.
The gas overlying the emission regions would itself serve as an absorber 
of the radio and X-ray radiation.
An estimate of $\dot M$ can be made from the XMM data. The mass loss rate is
a measure of the column depth $\int n dr$ rather than the
$\int n^2 dr$ which appears in the radio (free-free) absorption. 
The stellar wind would absorb the X-rays below 8 keV due to the
dominant photo-electric absorption of the metals.
The time at which $\tau \sim 1$ is reached for photons of 
energy $E_{\rm keV}$ is (\cite{Che94}):

$$ t_{\rm X} = C_{5} \dot M_{-5} u_{\rm w1}^{-1} u_4^{-1} E_{\rm keV}^{-8/3} $$

The value of the constant $C_5 (n)$ depends upon the density profile (n)
of the ejecta ($\rho_{\rm ej} = \rho_0 (t/t_0)^{-3} (u t/r)^n$). Using a 
typical value of the profile index $n=10.2$,
the epoch of X-ray detection as day 5 and $E_{keV}= 0.7$, we determine
$\dot M_{-5} / v_{\rm w1} \leq 6.5 \times 10^{-2}$. 
From optical
spectral measurements, the velocity of the pre-supernova wind
was $ \simeq 580$ km s$^{-1}$ (\cite{Leo02}). This
implies that $\dot M \leq 4 \times 10^{-5} M_{\odot}$. The above expression
assumes that the heavy elements contributing to the opacity are mostly
neutral -- this upper limit would be less stringent if Oxygen in the 
undisturbed circumstellar medium is heavily ionised, in which case the
X-ray absorption is reduced.

A similar mass loss rate can also be derived from the first radio detection
of SN 2002ap in the 1.4 GHz band on 2002 Feb 1.93 (i.e. on day 5) when
the optical depth in the radio frequency band would have been $\sim 1$.
At this time the dominant opacity source for radio radiation is
the free-free opacity from fully ionised wind (\cite{Che82}):
From the radio data, we estimate the upper limit on the mass loss rate as 
$\dot M \leq 6 \times 10^{-5} \rm M_{\odot} \; yr^{-1}$.

%
%

\section{Radiative processes in the early SN}
\label{sec: radiative}

Our spectral modeling of XMM observation of SN 2002ap shows that
both power-law and thermal bremsstrahlung models give acceptable
fits to early X-ray data. Since multiwaveband 
data from UV, optical and radio are also available from this 
supernova, these can
be additionally utilised to restrict models of prompt emission. 
Among the possible mechanisms invoked in the context of radio loud
X-ray supernovae are: 
a) direct synchrotron emission from radio to X-ray band,
b) free-free emission from the hot gas consisting
of swept up mass behind the blast wave (circumstellar) shock and that
behind the reverse shock, and
c) inverse Compton scattering of optical photons emitted by the supernova
by the hot electrons.

Case (a): As in the case of SN 1980K (\cite{Can82}), the direct synchrotron 
radiation 
mechanism for SN 2002ap prompt emission can be eliminated.
The power law spectrum in frequency for the radio flux observed from
SN 2002ap is: $F_{\nu} = 3 {\rm mJy} (3.2 \rm GHz /\nu)^{0.9} (t/day)^{-0.9}$
(unabsorbed flux from Berger et al., 2002). If the same ambient magnetic
field ($B \sim B_{eq} \sim 0.2$ Gauss) implied by equipartition arguments
were to hold, to generate 
X-ray energies (1 keV) by synchrotron mechanism, would imply the presence
of highly relativistic electrons ($\gamma \geq 1.4 \times 10^6$). 
On day 5, a straight extrapolation of the radio spectrum with the above
flux would imply a flux of only 58 picoJansky at 1 keV. 
Since the $90\%$ energy bandwidth for a power law distribution of energy 
index -1.6 in the 
XMM band is: 3.7 keV, this would imply an X-ray flux of about 
$5 \times 10^{-16} \rm erg \; s^{-1} cm^{-2}$. This is far less than what
is observed by XMM. Thus a direct radio to X-ray synchrotron radiation
mechanism is unviable, or produces an insignificant part of the observed
X-rays. 
Therefore, different populations of electrons appear to
be responsible for the radio and X-ray fluxes that are seen. 

Case (b): The circumstellar gas is heated to high temperatures 
($\geq 100$ keV) by the shock 
resulting from the piston-like expanding supernova envelope. 
%
This shocked gas cools by free-free emission and/or Compton cooling.
The free-free luminosity from 
the circumstellar (blast-wave) and reverse shocks
(\cite{Fra82}, \cite{Che01}) is:
$$ L_{\rm i} = 3\times 10^{39} g_{\rm ff} C_n \Big({\dot M_{-5}\over
u_{\rm w1}}\Big)^2 \Big({t\over 10 \ \rm day}\Big)^{-1} \rm erg \; s^{-1} $$
where for the reverse shock,
$C_n =(n-3)(n-4)^2/4(n-2) = 7.875$ for n=10 and $C_n = 1$ for the
circumstellar shock.  $g_{ff}$ is the Gaunt factor.

With $\dot M = 3.36 \times 10^{-5} M_{\odot} \rm yr^{-1}$ 
and with a wind velocity
$u_w = 1000$ km s$^{-1}$ (i.e. $u_{w1} = 100$), at t=5 days, the 
X-ray flux at 7.3 Mpc would be: $8.2 \times 10^{-15}$  erg cm$^{-2}$ s$^{-1}$.
This flux would have been near the observed flux that has been
detected from SN 2002ap by XMM. However, 
with the reported high ejecta velocities and the implied temperature
of the shocked ejecta and CSM along with the limited 
cool absorbing shell at this early stage, one would expect a flat tail
of high energy photons up to about 100 keV (\cite{Fra82}). 
Since the XMM spectrum 
on day 5 is not so hard (see Table 1 where
the thermal bremsstrahlung temperature is quite modest), the free free
emission is also an insignificant part of the {\it early}
radiative budget.
Thermal X-ray emission (i.e. bremsstrahlung) will however become more
dominant compared to the inverse Compton process (case c below)
as the supernova ages.

Case c): The dominant cooling and radiative mechanism at early stages
when the optical photospheric temperature 
$T_{eff} \geq 10^4 K$, is Compton cooling (\cite{Fra82}).
Optical photons from the photosphere can undergo repeated scattering off
the hot electrons in the shocked region (notably the shocked CSM)
and a power law 
photon distribution (in energy) can result even if the electron distribution
is not a power law, and even
if the electron scattering depth $\tau_{\rm e}$ remains small.
%
%
\cite{Poz77} reported Monte Carlo calculations
and their analytical approximations of the emergent Compton
scattered spectrum. The spectrum depends on the optical depth $\tau_{\rm e}$ and the 
temperature $T$ of the electron plasma.
For photons 
undergoing electron scattering with cross-section $\sigma_{T}$ 
in the shocked gas is: 
$$\tau_{\rm e}  = {\dot M \sigma_{\rm T} \over 4 \pi m_{\rm p} R_{\rm s} u_{\rm w}} 
\Big(1 - {R_{\rm opt} \over R_{\rm s}}\Big)$$
%
%
%
The flux density and the energy index $\gamma$ originating from
Comptonization are (\cite{Fra82}) :
$$ {\cal F}_{\nu}^{\rm Compton} \sim \tau_{\rm e} {\cal F}_{\nu}^{\rm opt} 
(\nu_o/\nu)^{\gamma} \rm erg \; s^{-1} cm^{-2} Hz^{-1} $$
with
$$ \gamma (\gamma + 3) = -{m_{\rm e}c^2\over kT_{\rm e}} \rm ln\Bigl[{\tau_{\rm e}\over
2}(0.9228- \rm ln \tau_{\rm e})
\Bigr] $$
Here, $\nu_{\rm o}$ is the peak frequency in the (optical) input spectrum,
and $F_{\rm o}$  the optical flux density. $R_{\rm opt}$ is the radius of photosphere.
$R_s$ is the position of the plasma element. 

We note that on day 5 the unabsorbed (or dereddened) X-ray
and optical flux densities derived from XMM and ground based
observations (Sect. \ref{sec: combined}) imply an optical 
to X-ray power law of $\gamma =2.5-2.8$
and a logarithm of the ratio of flux densities of $\approx 7.4$.
The optical to X-ray power law index is consistent with the 
XMM results
(see Table \ref{tab: EPIC-PN}). We determine the optical depth 
and temperature conditions of the Comptonizing plasma for $\gamma
=3 $ in Table 8. 
The presupernova scenarios typical for type Ic SNe, for which the
optical depths are reported in Table \ref{tab: progenitor} are discussed
in the next section.
The plasma
has the maximum optical depth at twice
the optical photosphere radius. The latter was taken
as $3.4 \times 10^{14} \rm \; cm$. Since most of the X-ray 
emission  would take place at $\sim \tau_{\rm max}$, the relevant
plasma outflows with a velocity  of approximately 
$16,000$ km s$^{-1}$.

\begin{table}
\caption{Comptonizing plasma properties at t = 5d 
with $\gamma=3 $ and $ R_{opt} = 3.4 \times 10^{14}$ cm
\label{tab:  progenitor}}
\begin{tabular}{ccccc}
\\
\hline
Scenario    & $\dot{M}_{-5}$  & $u_{w1}$ & $\tau_{e}$ &  $T_{e}$ \\
            & $10^{-5} \rm M_{\odot}/yr$ & 10 km/s &         &  $10^9$ K                   \\
\hline
Wolf-Rayet  & 1.5  & 58  & $4 \times 10^{-4}$ & $2$ \\
            & 3    & 100 & $4.4 \times 10^{-4}$ & $2$ \\
\hline
Interacting & 10 & 58 & $2.5\times 10^{-3}$ & $ 1.5$ \\
Binary      &    &    &                     &        \\                   
Case-BB     & 10 & 10 & $1.5 \times 10^{-2}$ & $1.1$ \\
 
\hline
\end{tabular}
\end{table}

\section{Discussion and Conclusion}
\label{sec: Discussion}

In this paper, we presented an analysis of the XMM data of the
SN 2002ap field and have obtained spectral model fits to the
prompt X-ray emission. We compare the X-ray image with the GMRT 610 MHz
radio image obtained three days apart. While we find no radio
counterpart of the SN at such low frequencies, several sources
in the field have radio and X-ray counterparts.
 
We compare the radio data obtained from three different supernovae
in their early phases and model these using the synchrotron self
absorption model. SN 1998bw(Ic)  with a GRB counterpart had very different
radiosphere radius and equipartition
angular sizes at approximately the same time in their evolution
compared to two other SNe: SN 2002ap(Ic)  and SN 1993J(IIb).
 
We model the early X-ray emission with inputs from optical photometry
and light curve and 
find that the inverse Compton scattering of optical photons from
the supernova photosphere by hot electron plasma can account for
the observed early X-ray flux and the spectrum for modest electron
temperatures and optical depths. Thermal processes are inefficient
initially and would be important only as the supernova ages.
 
Mass loss rates and stellar wind velocities of the
progenitor stars determine the optical depth of shock heated
matter due to electron scattering,
-- a key parameter in the production of the X-ray flux
from the lower energy photons. These in turn depend upon the
scenario and progenitor configuration giving rise to type
Ic SNe, e.g.:
 
1. Massive Wolf-Rayet(WR) stars which have lost their hydrogen and helium
envelope before the explosion (\cite{Lan99})
have an empirically determined
mass loss rate of $1.5 - 3.2 \times 10^{-5} M_{\odot}$
(after taking into account
effects due to clumpy medium, (\cite{Ham98}, \cite{Wil98}) and terminal
wind velocities $\sim 1000$ km s$^{-1}$ depending
upon the type of the WR star.

2. Interacting binaries - in particular, in a case BB mass transfer
from a helium star overflowing its Roche-lobe to a companion
removes most of the helium rich layers before
the type Ic SN.
Habets (1985) finds
that a 2.5 $M_{\odot}$ helium star 
during carbon shell burning stage expands to
a red giant dimension of 18 $R_{\odot}$ and sustains an
average mass transfer rate of
$10^{-4} \rm M_{\odot} yr^{-1}$
lasting about 3000 years.
During this time, the terminal wind speed,
of the mass losing
star would be typically $100$ km s$^{-1}$.
 
 
Using the relevant parameters in the above two scenarios, we derive
electron optical depth encountered by the intermediate energy
photons as listed in the Table \ref{tab: progenitor}.
It is evident that both ranges of electron optical depths
for the Compton boosting process 
remain viable
alternatives. For the case of the interacting binary model
where the optical depths are somewhat larger, the implied
electron temperatures required for the plasma would be lower
than in the single star WR model. Such temperatures are well
within the range expected for hot circumstellar gas
($T_{cs} \geq 3.6 \times 10^9 K$)
even for modest velocities of $16,000 - 20,000$ km s$^{-1}$
for the hot electron plasma moving above the optical photosphere.
\begin{acknowledgements}
This work made use of XMM Target of Opportunity data.
F. K. Sutaria would like to thank the XMM-Newton
helpdesk for many useful communications during the pipeline processing
of raw data. We acknowledge the use of Chandra archival data
on M74.
We thank the staff of the GMRT that made the radio observations possible. 
We also thank the NRAO staff for providing the Astronomical Image 
Processing Software (AIPS) for radio data analysis.
\end{acknowledgements}

{}


\end{document}